\documentclass[11pt]{iopart}
\usepackage{graphicx}
\usepackage[english]{babel}
\usepackage{amsfonts}
\usepackage{amssymb}
\expandafter\let\csname equation*\endcsname\relax
\expandafter\let\csname endequation*\endcsname\relax
\usepackage{amsmath}
\usepackage[ansinew]{inputenc}
\usepackage{mathtools}
\usepackage{wasysym}
\usepackage{color}
\usepackage{verbatim}
\usepackage{bm}
\usepackage{bm}
\usepackage[b]{esvect} % Vector notation 
\usepackage{braket}
\usepackage{textcomp} % text companion fonts
\usepackage{microtype} % makes pdf looks better
\usepackage{hyperref}
\hypersetup{colorlinks=true}
\usepackage[all]{hypcap} % let hyperlinks correctly point to figures 

\newcommand{\akd}{a^{\dagger}_{k}}
\newcommand{\ak}{a^{\phantom{\dagger}}_{k}}
\newcommand{\w}{\omega}

\newcommand{\ba}{\begin{array}}
\newcommand{\ea}{\end{array}}
\newcommand{\bit}{\begin{itemize}}
\newcommand{\eit}{\end{itemize}}
\newcommand{\Le}{\left}
\newcommand{\Ri}{\right}
\newcommand{\nn}{\nonumber}

\newcommand{\f}{\frac}

\newcommand{\mrm}{\mathrm}

\newcommand{\ra}{\rangle}
\newcommand{\la}{\langle}
\newcommand{\ua}{\uparrow}
\newcommand{\da}{\downarrow}
\newcommand{\ii}{i}
\newcommand*\conj[1]{{#1^*}}
\newcommand*\conjk[1]{{#1^{k*}}}

\begin{document}

\title{
Spontaneous emission of Schr\"odinger cats in a waveguide at ultrastrong
coupling}

\author{Nicolas Gheeraert}
\address{Institut N\'{e}el, CNRS and Universit\'e Grenoble Alpes, F-38042 Grenoble, France}
\author{Soumya Bera}
\address{Max-Planck-Institut f\"ur Physik Komplexer Systeme, 01187 Dresden,
Germany}
\address{Institut N\'{e}el, CNRS and Universit\'e Grenoble Alpes, F-38042 Grenoble, France}
\author{Serge Florens}
\address{Institut N\'{e}el, CNRS and Universit\'e Grenoble Alpes, F-38042 Grenoble, France}

\begin{abstract}
Josephson circuits provide a realistic physical setup where the light-matter fine 
structure constant can become of order one, allowing to reach a regime dominated by 
non-perturbative effects beyond standard quantum optics. 
Simple processes, such as spontaneous emission, thus acquire a many-body
character, that can be tackled using a new description of the time-dependent 
state vector in terms of quantum-superposed coherent states.
We find that spontaneous atomic decay at ultrastrong coupling leads to the emission 
of spectrally broad Schr\"odinger cats rather than of monochromatic single photons. 
These cats states remain partially entangled with the emitter at intermediate stages
of the dynamics, even after emission, due to a large separation in time scales between 
fast energy relaxation and exponentially slow decoherence. Once decoherence 
of the qubit is finally established, quantum information is completely transfered to
the state of the emitted cat.
\end{abstract}

\date{\today}

\maketitle

{\it Introduction.}
Photons describe the granular structure of the electromagnetic field 
radiated by single coherent sources, such as atoms and quantum dots. These 
quanta of light constitute well-defined monochromatic excitations 
because the spontaneous emission rate $\Gamma$ is much
smaller than the transition frequency $\Delta$ of an emitter. More 
precisely, for atomic decay in three-dimensional space~\cite{Sargent}, the
width of the transition lines, $\alpha=\Gamma/\Delta$, also defines a 
dimensionless coupling constant $\alpha =(\mathcal{P}/e\lambda)^2\alpha_\mathrm{QED}$ 
involving two small factors: the ratio of the atomic dipole
$\mathcal{P}$ to the photon wavelength $\lambda$, and the vacuum fine structure 
constant $\alpha_\mathrm{QED}\simeq 1/137$.
Typically $\mathcal{P}/e\lambda\lesssim \alpha_\mathrm{QED}$, so that the 
light-matter coupling to a 3D continuum,
$\alpha\lesssim(\alpha_\mathrm{QED})^3\simeq 10^{-6}$, is vanishingly small.

With the advent of microwave quantum optics in superconducting
one-dimensional waveguides~\cite{Schoelkopf,Astafiev,Abdumalikov,Hoi1,Hoi2,vanLoo,Houck,Haeberlein}, 
the coupling constant can reach the still small value 
$\alpha\simeq\alpha_\mathrm{QED}\simeq10^{-2}$.
Thus, a coupling of order one can only be attained by tweaking 
the fine structure constant itself. This requires the use of
superconducting waveguides with high impedance, since 
$\alpha_\mathrm{QED}=Z_\mathrm{vac.}/2R_\mathrm{K}$ 
can be interpreted as the ratio of the vacuum impedance 
$Z_\mathrm{vac.}=1/\epsilon_0 c\simeq 376\; \Omega$ 
to the quantum of resistance $R_\mathrm{K}=e^2/h\simeq25812\;\Omega$.
Long chains of Josephson junctions~\cite{Masluk,Altimiras,Bell,Weissl} constitute 
low-loss metamaterials for the propagation of microwave photons with characteristic 
impedance $Z$ up to the order of $R_\mathrm{K}$, allowing to obtain values of the 
coupling constant $\alpha=Z/R_K$ of order one. This physical regime reveals anomalous 
scattering properties of 
photons~\cite{LeHur,Goldstein,Sanchez,Peropadre,Gheeraert} impinging on a non-trivial 
dressed vacuum~\cite{Snyman1}. This Letter aims at answering a seemingly
simple question: Is the radiation from a single emitter still described by a
discrete photon in the ultrastrong coupling regime? 
%is the one-photon state still well defined as quantum of 
%radiation from a single emitter in the ultrastrong coupling regime?

\begin{figure*}[th]
\includegraphics[width=0.99\linewidth]{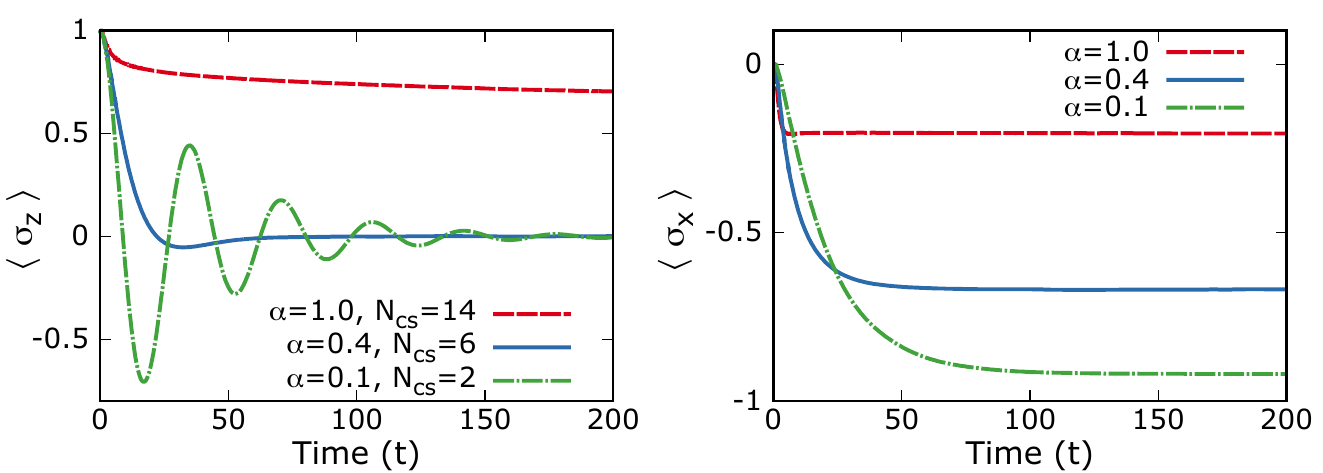}
\caption{(Color online) Left panel: decoherence process $\big<\sigma_z(t)\big>$
with typical decay time $T_2$, for increasing dimensionless coupling $\alpha=0.1, 0.4, 1.0$ 
(with coherent state number $N_\mathrm{cs}=2,6,14$, as required to reach convergence in the 
respective regimes). In all computations, $\Delta/\omega_\mathrm{p}=0.2$ and $N_\mathrm{modes}=800$.
Right panel: energy relaxation process $\big<\sigma_x(t)\big>$ with typical decay time 
$T_1$, for the same parameters. 
For $\alpha\ll1$, decoherence and relaxation times fulfill the usual relation 
$T_2=2T_1$, but for $\alpha=1.0$ the two time scales are widely different, with $T_2\gg
T_1$, see text for a physical discussion of this effect.}
\label{Decay}
\end{figure*}

Clearly, the radiation released for $\alpha\simeq1$ is strongly
non-monochromatic, because ultrafast energy relaxation processes gives
a large broadening of the atomic transition lines.
In addition to modified spectral properties, drastic effects also occur 
in the quantum statistics of the radiation. Indeed, in contrast 
to spontaneous decay at weak coupling, where the photon state emerges as 
an $n=1$ Fock excitation~\cite{Sargent}, the radiated field at 
ultrastrong coupling contains states with photon number larger 
than one. This surprising result can be understood as follows: photons with energy 
much smaller or much larger than the atomic transition frequency can be excited 
because one is not limited to resonant transitions in the non-perturbative
regime of quantum electrodynamics (QED). But for low energy photons,
the large available energy and the strong coupling constant make favorable the 
creation of multiple photon excitations. Infact, we find that the full quantum 
state of the electromagnetic field possesses a simple structure given by an 
optical Schr\"odinger cat, namely a superposition of strongly displaced coherent states. 
%This can be rationalized by turning the tables: at ultrastrong 
%coupling, the atom plays the role of an environment for its surrounding 
%electromagnetic field. This leads to the selection of the most robust quantum states, 
%the so-called pointer states~\cite{Zurek}, which are well described by coherent states
%in the context of quantum electrodynamics. 
We stress that cat states obtained by spontaneous decay in a large impedance environment 
are very different from the cats obtained in cavities~\cite{Haroche}, for two reasons. 
First, as already mentioned above, the radiation is spectrally broad at ultrastrong 
coupling, and thus the resulting cat states are rather localized in the time domain 
rather than in frequency. 
Secondly, we find that these cat states display an intrinsic loss of coherence
at longer times than the sudden time scale $T_1$ associated to energy relaxation, but 
shorter than the long decoherence time $T_2$ for complete memory loss of the initial 
atomic state. The separation of time scales $T_2\gg T_1$ at ultrastrong
coupling, see Fig.~\ref{Decay}, illustrates again the stark difference 
with the weak-coupling regime of quantum optics, where $T_2=2 T_1$, and will
be one of the main focus of the rest of the paper.

{\it Methodology.}
Having presented the physics at play, we must emphasize that the computational 
aspects of quantum electrodynamics at ultrastrong coupling are far from trivial, due to
the breakdown of perturbation theory~\cite{Leggett,LeHurReview,Snyman1}.
In contrast to cavity-QED at ultrastrong coupling~\cite{Devoret,Niemczyk,Forn},
the Hilbert space is unmanageable for high impedance waveguides because: 
i) a large number of electromagnetic modes are involved, 
up to $N_\mathrm{modes}=1000$ for long chains of Josephson junctions;
ii) the average total number of photons $\bar n_\mathrm{tot}$ is larger than one.
For the case $\alpha=1.0$ considered in the following, one obtains up to
$\bar n_\mathrm{tot}=6$ photons, so that the full quantum mechanical problem requires to tackle 
more than $(N_\mathrm{modes})^{\bar n_\mathrm{tot}}\simeq 10^{18}$ quantum states, 
well beyond the reach of brute force diagonalization. Based on the
physical idea that coherent states are the most stable quantum states at ultrastrong 
coupling~\cite{Bera1,Bera2,Snyman2}, we propose here a new methodology for harnessing 
cat states which relies on an efficient computational algorithm. This technique is not 
only conceptually simpler than state of the art 
methods~\cite{WangThoss,BullaAnders,Hofstetter,Zueco,Chin}, but also very powerful.

We will specify our study to the standard model of waveguide-QED, described by 
the Hamiltonian:
\begin{equation}
\label{hamiltonian}
H = \f{\Delta}{2} \sigma_x + \sum_k \omega_k a_k^\dag a_k - \sigma_z \sum_k
\f{g_k}{2} \bigl( a_k^\dag + a_k \bigr),
\end{equation}
defining the light-matter coupling $g_k$ of a given mode with frequency $\w_k$
to the two-level atom, described by Pauli matrices $\sigma_j$ and 
transition frequency $\Delta$. 
At variance with quantum optics conventions, we have intentionally written 
the atomic splitting as a $\sigma_x$ term, in order to emphasize the natural selection
of coherent states caused by the $\sigma_z$ light-matter coupling. Hence, the bare 
atomic ground state is $|g\big>=[|\uparrow\big>-|\downarrow\big>]/\sqrt{2}$ in our 
notation. The dimensionless coupling strength $\alpha$ can be encapsulated 
from the spectral density $J(\w) = \pi \sum_k g_k^2 \delta(\w-\w_k) = 2\pi\alpha \w 
e^{-\w/\w_\mathrm{p}}$, with $\w_\mathrm{p}$ the plasma frequency.
We assume here a linear dispersion relation $\w_k = k$, with unit speed of light in 
the medium, which is justified for atomic transitions well below $\omega_\mathrm{p}$.
In addition, we do not consider here the important effect of coupling the
high impedance finite-size Josephson waveguide to 50 $\Omega$ lines, which will restrict 
the spectrum to a set of discrete resonances (slightly broadened by the contacts). 
However, for a long-enough Josephson array, the limits in resolution will be well-below 
the characteristic scales of the system. Finally, we consider purely unitary dynamics, 
assuming that extrinsic losses, apart from the coupling to external contacts, are negligible.

As discussed above, the state vector is represented at all times 
by superposed coherent states~\cite{Bera1,Bera2,Snyman2}:
\begin{equation}
\label{Psi}
|\Psi(t)\ra = \sum_{m=1}^{N_\mathrm{cs}} 
\Big[ p_m(t) |f_m(t) \ra | \ua \ra + q_m(t) |h_m(t) \ra | \da \ra 
\Big].
\end{equation}
Here the set of amplitudes $\{p_m,q_m\}$ are complex and time dependent, and 
a set of discrete multimode coherent states are introduced: 
$|f_m(t)\big> = e^{\sum_{k=1}^{N_\mathrm{modes}}
[f_m(k,t) \akd - f_m^*(k,t) \ak]} |0\big>$. This decomposition allows in principle 
to target an arbitrary state of the full Hilbert space (with an
exponential cost). One must note a strong difference here with the standard
Glauber-Sudarshan decomposition, which relies on a continuous expansion
of the state vector onto coherent states. Because of the large number of modes
involved in waveguide QED at ultrastrong coupling, the continuous representation is 
not suitable for numerical purposes. Our discrete expansion~(\ref{Psi}) can
however be understood as a discretized version of the continuous integral
representation of an arbitrary wavefunction onto coherent states.
Most importantly, we argue in this paper that physical states obtained from 
standard protocols (such as spontaneous emission) can be efficiently simulated 
with a discrete set of coherent states, showing only a polynomial cost in the 
number of modes $N_\mathrm{modes}$ and coherent states $N_\mathrm{cs}$. 
This computational gain was previously demonstrated 
for the full ground state of~(\ref{hamiltonian}), and is here extended to the 
dynamics, allowing to simulate for the first time large systems, up to thousands 
of electromagnetic modes.

\begin{figure}[th]
\includegraphics[width=1.0\linewidth]{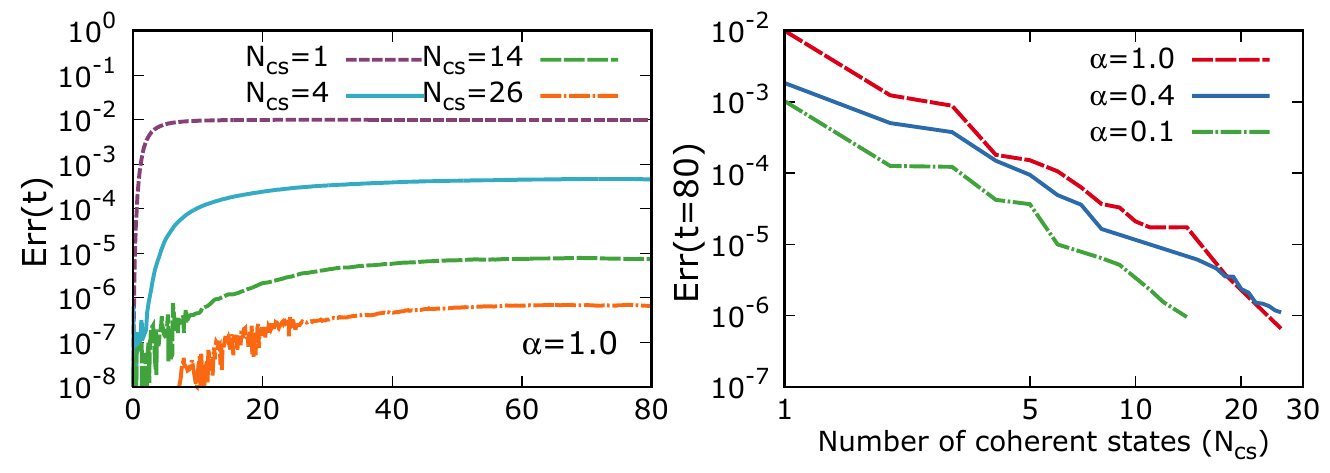}
\caption{(Color online) 
Left panel: error $\mathrm{Err}(t)$ as a function of time with respect to the exact Schr\"odinger
dynamics for the challenging case $\alpha=1.0$ (here the atomic splitting is
$\Delta/\omega_c=0.2$), which drops rapidly to 
zero with increasing $N_\mathrm{cs}$.
Right panel: error $\mathrm{Err}(t=T)$ at a fixed time $T=80$, as a function of coherent 
state numbers for $\alpha=0.1$, $\alpha=0.4$ (intermediate ultracoupling regime) and 
$\alpha=1.0$ (deep ultrastrong coupling regime), showing a scaling
$[N_\mathrm{cs}]^{-2}$.}
\label{Error}
\end{figure}

Technically, the exact Schr\"odinger dynamics controlled by
the Hamiltonian~(\ref{hamiltonian}) can be obtained from the real Lagrangian density:
\begin{equation}
\mathcal{L}= \big<\Psi(t)|\frac{i}{2} \overrightarrow{\partial_t} 
-\frac{i}{2} \overleftarrow{\partial_t} 
- \mathcal{H}|\Psi(t)\big>,
\end{equation}
by applying the time-dependent variational 
principle~\cite{Saraceno}, $\delta \int \mathrm{d}t \mathcal{L}=0$, upon arbitrary 
variations of the state vector~(\ref{Psi}).
This results in Euler-Lagrange equations $\frac{d}{dt} \frac{\partial \mathcal{L}}{\partial \dot{v}} =
\frac{\partial \mathcal{L}}{\partial v}$ for the set of variables 
$v=\{p_m,q_m,f_m(k),h_m(k)\}$, which can be solved by numerical
integration~\cite{Burghardt,Zhao,ZhaoMulti,Bera3} using a specially devised
algorithm (see Appendix for details).

We provide here clear proof of the good convergence of our algorithm.
First, the left panel of Fig.~\ref{Error} shows how the time-dependent
error $\mathrm{Err(t)}$ vanishes with the number of coherent states, 
for the deep ultracoupling regime $\alpha=1.0$. Here, the error is defined 
by the squared norm $\mathrm{Err}(t)\equiv \big<\Phi(t)|\Phi(t)\big>$ 
of the auxiliary state 
$|\Phi(t)\big>\equiv(i\partial_t -H)|\Psi(t)\big>$, which is zero for the exact
Schr\"odinger dynamics. 
Clearly the error is already small at all times for a single coherent state
$N_\mathrm{cs}=1$, and goes quickly to zero with increasing number of
terms in the decomposition Eq.~(\ref{Psi}). The precise scaling of the
algorithm with $N_\mathrm{cs}$ is demonstrated in the right panel of 
Fig.~\ref{Error}. We find that the error (here computed at a finite and
fixed time $T=80$) decreases typically with an inverse square power 
$[N_\mathrm{cs}]^{-2}$, independently of the coupling strength $\alpha$.
This shows that our methodology is based on a physically well-motivated 
decomposition of the state vector, and is not tied to a particular regime 
of the spin-boson model. The coherent state expansion of the time-dependent
state vector~(\ref{Psi}) thus provides numerically accurate results in 
all regimes of coupling for a small computational effort.

{\it Decoherence vs. relaxation times.} We consider first the important issue of the
time scales governing the physics in the ultrastrong coupling regime. For
this purpose, we prepare the initial state as $|\Psi(t=0)\big> =
|0\big>\otimes|\uparrow\big> = |0\big>\otimes
[|g\big>+|e\big>]/\sqrt{2}$ in a superposition of the two bare atomic levels,
and with the environment in its vacuum. 
Since the two-level splitting is described by a $\sigma_x$ coupling in 
Eq.~(\ref{hamiltonian}), one expects precession and decay to zero of the transverse 
spin component $\big<\sigma_z(t)\big>$ on a time scale $T_2$, and relaxation of the
longitudinal term $\big<\sigma_x(t)\big>$ towards its finite equilibrium value on
a time scale $T_1$. This standard behavior is well obeyed in the weak-coupling
regime $\alpha\ll1$, with $T_2=2T_1$, as seen from the dot-dashed green curve in the first 
two panels of Fig.~\ref{Decay}.
% Here the atom decays from the initial value $\big<\sigma_x(0)\big>=1$ to a final 
% stationary state with $\big<\sigma_x(\infty)\big>\simeq-0.9$.
% In contrast to the $\alpha\to 0$ quantum optics regime, the long-time 
% atomic occupation turns out to be different from -1, because the atom
% experiences a photonic dressing due to large counter-rotating terms.
% This renormalization effect is even more pronounced at larger dissipation 
% strength~\cite{LeHur,Bera1}, as seen by the curves for $\alpha=0.4,0.8$ in
% the upper panel of Fig.~\ref{Decay}.
For the intermediate value $\alpha=0.4$ (full blue curve), precession of 
$\big<\sigma_z(t)\big>$ is nearly overdamped, as is well 
established~\cite{Leggett}. 

Remarkably, the ultrastrong coupling regime $\alpha=1.0$ (dashed red curve) shows a striking 
decoupling between the decoherence time $T_2$ and the energy relaxation time $T_1$, 
with $T_2\gg T_1$. The underlying physics can be anticipated: energy relaxation,
related to emission of radiation, occurs on the short time scale $T_1$, because 
the excited atomic level is strongly damped at large $\alpha$. This is clearly seen
from the rapid saturation of $\big<\sigma_x(t)\big>$ in the middle panel of
Fig.~\ref{Decay} at $\alpha=1.0$. However, rapid decoherence is prohibited because 
the atom is dressed by its electromagnetic environment~\cite{Snyman1} on a large 
spatial scale $L_K = (\w_\mathrm{p}/\Delta)^{\alpha/(\alpha_c-\alpha)}$ 
(here $\alpha_c=1+\Delta/\omega_\mathrm{p}$, corresponding to the threshold of full 
localization). Decoherence thus takes a considerable time $T_2\simeq L_K$
for the complete relaxation of all quantum correlations between the dressed 
atom and the is larger than one,radiated field. This effect is seen by the very slow decay
of $\big<\sigma_z(t)\big>$ (dashed red curve in the left panel of 
Fig.~\ref{Decay}). Indeed, for $\alpha=1.0$ and $\Delta/\omega_\mathrm{p}=0.2$,
 $T_2\simeq3000$, which was not reached at the final time of our simulations.
As we will see in the following, this separation of time scales is key to understanding the 
physics at play, due to its strong impact on the structure of the emitted light.

\begin{figure}[th]
\includegraphics[width=0.99\linewidth]{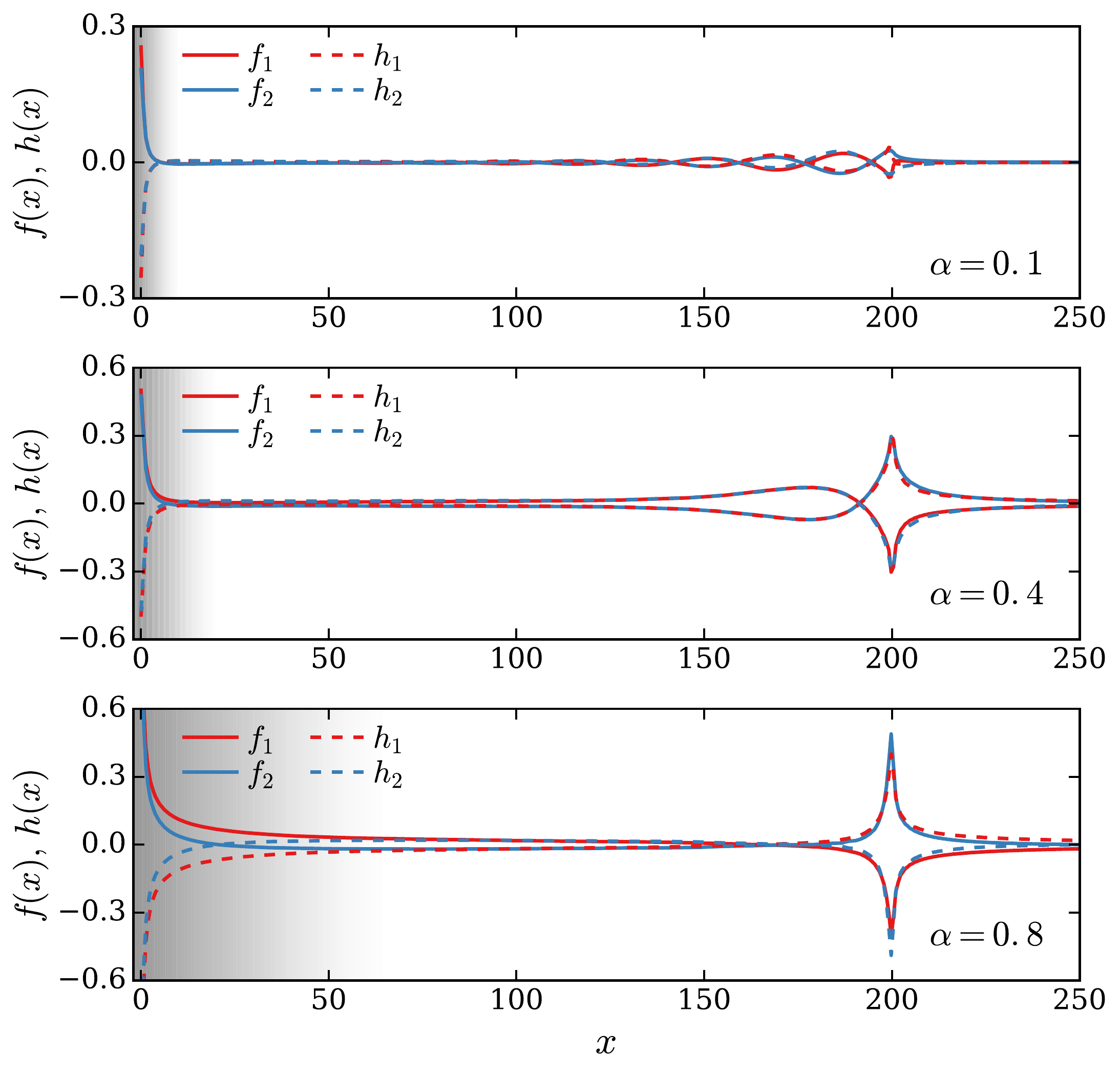}
\caption{(Color online) Real part of the coherent state amplitudes in real space
$f_m(x,T)$ and $h_m(x,T)$ (for $m=1,2$) at a 
time $T=200$ long enough that the emitted wavepacket is uncorrelated from 
the entanglement cloud associated to the dressed atomic ground state (denoted 
as a shaded area). Here $\Delta/\omega_\mathrm{p}=0.2$, and $N_\mathrm{cs}=2$, with $\alpha=0.1$ 
(top panel), $\alpha=0.4$ (middle panel), and $\alpha=0.8$ (bottom panel).}
\label{Displacement}
\end{figure}
{\it Quantum states of spontaneous emission.} 
We turn to the emission protocol, taking now the bare excited level of the atom 
$|\Psi(t=0)\big> = 
|0\big>\otimes[|\uparrow\big>+|\downarrow\big>]/\sqrt{2}
=|0\big>\otimes|e\big>$ as the initial state.
%The system again evolves unitarily at later times according to 
%Hamiltonian~(\ref{hamiltonian}). 
According to Wigner-Weisskopf theory~\cite{Sargent} valid at 
weak coupling $\alpha\ll1$, one expects 
spontaneous emission of a single photon and decay of the atom 
towards its bare ground state. Although our theory~(\ref{Psi})
is based on coherent states and not Fock excitations, we can show that
it does recover the standard quantum optics results at $\alpha\to0$,
while providing a simple physical picture in the ultrastrong coupling regime.
For this purpose, we plot in Fig.~\ref{Displacement} the real part 
of the coherent state amplitudes $f_m(x,T)$ and $h_m(x,T)$ (for $m=1,2$) 
for $N_\mathrm{cs}=2$, as a function of the spatial separation $x$ along
the waveguide (the atom is located at $x=0$).
Here we choose a time $T\gg T_2\simeq L_K$ long enough that the system has fully relaxed to its 
dressed ground state. 
% Due to the obvious $\mathbb{Z}_2$ symmetry (transforming 
% $a^\dagger\to-a^\dagger$ and $\sigma_z\to-\sigma_z$) of
% Hamiltonian~(\ref{hamiltonian}) that is also preserved by our initial 
% wavefunction $|\Psi(0)\big>$, we have the exact symmetry relations 
% $f_m(x,t)=-h_m(x,t)$ and $p_m(t)=q_m(t)$ for all $n$ at all times.
% In order to ease the physical interpretation of these curves, let us first
% discuss the short distance behavior of the coherent state amplitudes. For $x<L_K$, 
The emitted wavepacket is clearly seen at a distance $x\simeq T$, due to ballistic
propagation of the wavefront, and the displacements satisfy the following relations: 
$f_1(x,t)\simeq -f_2(x,t) \simeq -h_1(x,t)\simeq h_2(x,t)\equiv f_\mathrm{wp}(x-t)$ 
[in addition, $p_1=-p_2=q_1=-q_2\equiv p$ (not shown)].
Within the spatial region $x<L_K$ where the atom is dressed by its static 
surrounding cloud, a different set of relations is observed: 
$f_1(x,t)\simeq f_2(x,t) \simeq -h_1(x,t)\simeq -h_2(x,t) \equiv 
f_\mathrm{cl}(x)$.
Factoring each coherent state in real space $|f_1\ra = |f_\mathrm{cl}\ra\otimes
|f_\mathrm{wp}\ra $ simplifies the wavefunction~(\ref{Psi}):
\begin{eqnarray}
\label{factorize}
\hspace{-1.2cm} |\Psi(T)\ra &\simeq& \big[p_1 |f_1\ra + p_2 |f_2\ra \big] |\ua\ra 
+ \big[q_1 |h_1\ra + q_2 |h_2\ra \big] |\da\ra \\
\nonumber
&\simeq&
\big[p |f_\mathrm{cl}\ra |f_\mathrm{wp}\ra 
- p |f_\mathrm{cl}\ra |-f_\mathrm{wp}\ra \big] |\ua\ra 
+ \big[p |-f_\mathrm{cl}\ra |-f_\mathrm{wp}\ra 
- p |-f_\mathrm{cl}\ra |f_\mathrm{wp}\ra \big] |\da\ra \\
\nonumber
&\simeq&
\big[|f_\mathrm{cl}\ra |\ua\ra - |-f_\mathrm{cl}\ra
|\da\ra\big]\otimes p
\big[|f_\mathrm{wp}\ra - |-f_\mathrm{wp}\ra \big].
\end{eqnarray}
The last line in~(\ref{factorize}) is straightforwardly interpreted as 
the absence of correlations between the dressed qubit 
$|\Psi_\mathrm{cl}\ra \equiv \big[|f_\mathrm{cl}\ra |\ua\ra - |-f_\mathrm{cl}\ra
|\da\ra\big]/\sqrt{2}$ (which comprises both the atom and its entangled
neighboring cloud in the waveguide) and the emitted wavepacket $|\Psi_\mathrm{wp}\ra \equiv 
\sqrt{2}p\big[ |f_\mathrm{wp}\ra - |-f_\mathrm{wp}\ra\big]$.
This is physically expected as the system relaxes at long times to a unique
dressed ground state, while emitting a stream of electromagnetic radiation 
carrying the excess energy but no quantum correlations with the atom.
We stress that the above expressions for $|\Psi_\mathrm{cl}\ra$ and $|\Psi_\mathrm{wp}\ra$
are only approximate, as small quantum corrections arise at increasing 
$\alpha$~\cite{Bera1,Bera2}, which are accounted for by the terms
$n>2$ in the expansion~(\ref{Psi}).

The approximate wavefunction~(\ref{factorize}) nicely recovers the result of
Wigner-Weisskopf theory in the quantum optics regime $\alpha\to 0$. Indeed, in this 
case the displacements in the dressing cloud are vanishingly small, $|f_\mathrm{cl}(x)|\ll1$,
and thus $|\Psi_\mathrm{cl}\ra = |0\ra \big[|\ua\ra - |\da\ra\big]/\sqrt{2} \equiv |0\ra
|g\ra$, so that the atom has correctly relaxed to its bare ground state. The
quantum state describing the emitted light also simplifies, since the 
displacements in the wavepacket are small, $|f_\mathrm{wp}|\ll1$. A
first-order Taylor expansion of the coherent state gives $|\Psi_\mathrm{wp}\ra = 2 \sqrt{2}p \sum_k
f_\mathrm{wp}(k) e^{ikt} a_k^\dagger|0\ra$, which is the expected one-photon Fock state.
The monochromatic nature of the emitted light can be seen
from the spatiotemporal oscillations of the fields in the upper panel of
Fig.~\ref{Displacement}, with an envelope controlled by the underdamped dynamics 
of the qubit (first panel of Fig.~\ref{Decay} for $\alpha=0.1$).
Moving towards the ultrastrong coupling regime for increasing $\alpha$ values
(middle and lower panel in Fig.~\ref{Displacement}), two major changes occur.
First, the displacements in the entanglement cloud $x<L_K$ penetrate deeper and
deeper within the waveguide, due to the increase of the screening length $L_K$
with $\alpha$, as shown by the shaded area in Fig.~\ref{Displacement}. Second,
the emitted wavepacket becomes very localized temporally, with associated
displacements that clearly grow in magnitude. This indicates that the radiation
is spectrally broad, and that the number of emitted photons grows with
increasing $\alpha$, as was anticipated.

\begin{figure}[th]
\includegraphics[width=0.99\linewidth]{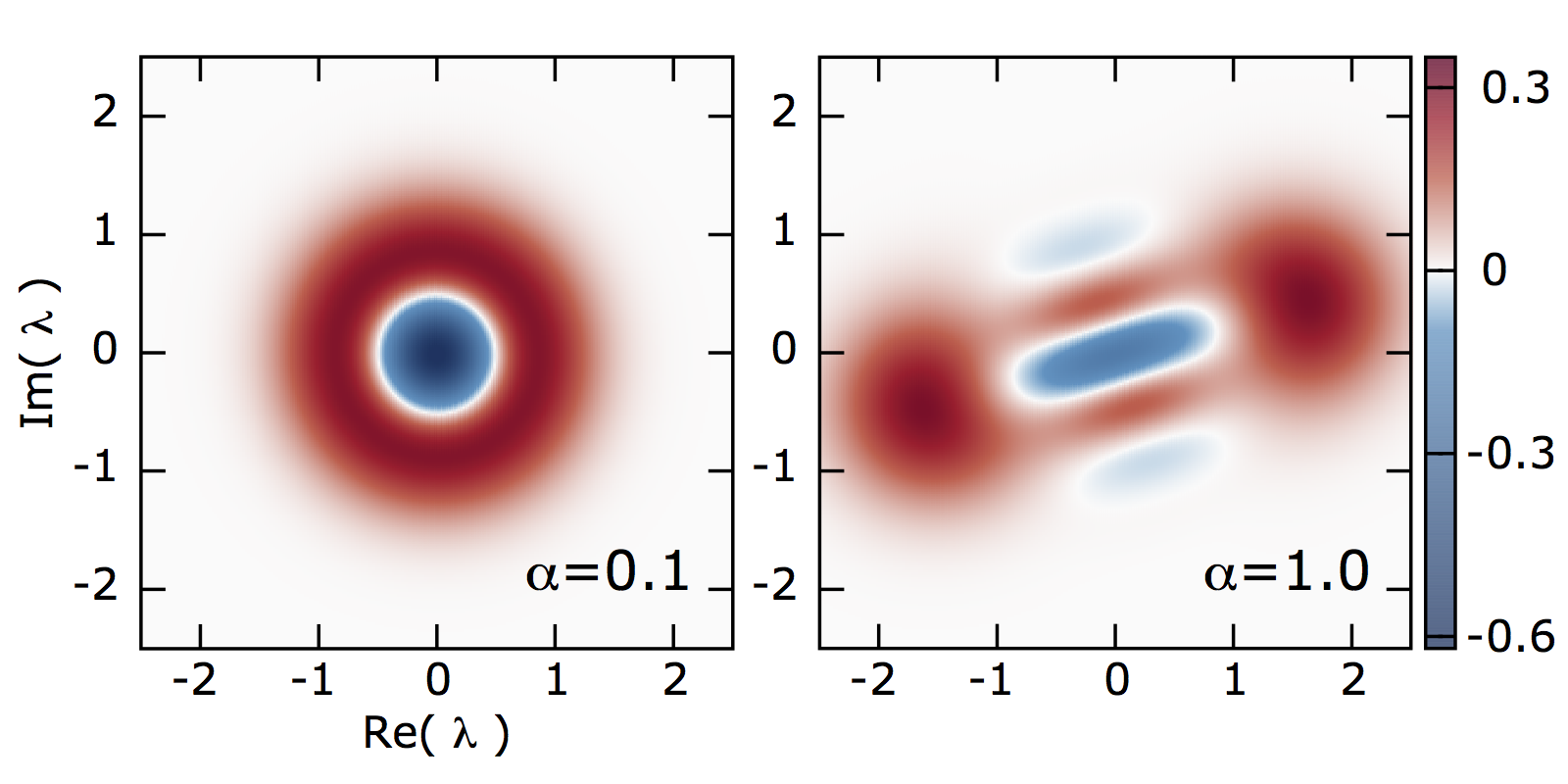}
\caption{(Color online) Wigner distribution $W(\lambda)$ at time $T=900$
identifying a one-photon state at $\alpha=0.1$ (left panel) and a partially 
coherent cat state at $\alpha=1$ (right panel) with two positive 
classical lobes and a negative region with
reduced amplitude compared to $-2/\pi$ for a fully coherent cat. The 
function was computed with $N_{\mathrm{cs}}=14$ and $N_{\mathrm{modes}}=1800$.}
\label{Wigner}
\end{figure}
{\it Nature of spontaneously emitted cats.}
Let us finally analyse in more detail the quantum properties of 
the emitted radiation, starting with the approximate expression 
$|\Psi_\mathrm{wp}\ra = \sqrt{2}p\big[ |f_\mathrm{wp}\ra -
|-f_\mathrm{wp}\ra\big]$ for the emitted wavepacket. Although this state is a
Fock state when the average number of emitted photons is close to one, as shown
above, it clearly turns into an odd parity Schr\"odinger cat when the 
displacements grow large, as seen for $\alpha=0.8$ in the lower panel in 
Fig.~\ref{Displacement}.
In order to check this idea more precisely, we compute the Wigner distribution
of the emitted wavepacket.
In contrast to photons released into a cavity~\cite{Deleglise,Haroche}, we stress
again that the radiation is not purely monochromatic, due to the significant damping effect 
on the atom caused by the strong coupling to the waveguide. Accordingly, following standard 
practice~\cite{Eichler}, one uses an optimized temporal filter $w(t)$ of the output
signal. 
The filter function $w(t)$ is defined as to match precisely the shape
of any of the wavefronts in Fig.~\ref{Displacement} (which show the same
amplitude), in the spatial domain where the emitted signal is decoupled from 
the short distance cloud (here in the approximate range $100<x<250$).
For the values of $\alpha\leq0.8$ in Fig.~\ref{Displacement}, a measurement
time $T=200$ is sufficient to ensure this decoupling, but for the computation of
Fig.~\ref{Wigner} with $\alpha=1.0$, a longer time $T=900$ was required, due
to the increase in $T_2$.

Owing to the linear dispersion in the waveguide, the averaging can be
performed spatially by defining an effective creation operator $b^\dagger = \sum_x 
w(x) a^\dagger(x) \Theta(x-L_K)$, in which a $\Theta$-function is used to take out the 
static bound component of the screening cloud. The global scale of the
filter function $w(t)$ is set by imposing standard commutation relation for the
effective mode, namely $[b,b^\dagger]=1$.
The Wigner distribution~\cite{Haroche} is then computed using the standard 
expression $W(\lambda) = \int (\mathrm{d}^2\beta /\pi^2) C_s(\beta) 
e^{\lambda\beta^*-\lambda^*\beta}$, with the symmetrized correlation function
$C_s(\beta)=\big<\Psi|e^{\beta b^\dagger-\beta^* b} |\Psi\big>$.
For $\alpha=0.1$, the phase space distribution in the left panel of Fig.~\ref{Wigner} 
shows the characteristic circular form of the $n=1$ Fock state with negative 
quasi-probability $-2/\pi$ at the origin.

At ultrastrong coupling for $\alpha=1.0$, the emitted radiation
undergoes radical changes, as shown in the right panel of Fig.~\ref{Wigner}.  
The Wigner distribution now presents two positive lobes (signature
of the two classical configurations of the cat), but also a negative region 
near the origin, fingerprint of the characteristic quantum interference, 
or ``whiskers'', of a Schr\"odinger cat.
Surprisingly, the maximum negative amplitude does not reach the expected value
$-2/\pi$ of a perfectly coherent cat, although our system does not present any
extrinsic source of decoherence for the optical modes (such as leaks into a 3D
continuum). We argue that the physical source of decoherence is
the dressed qubit itself, a very unusual feature.
This phenomenon can be understood from the qubit dynamics shown in Fig.~\ref{Decay}, 
in relation to the separation of time scales $T_2\gg T_1$. Indeed, the cat state
is emitted on a short scale $T_1\simeq 1/\w_\mathrm{p}$ in the ultrastrong
coupling regime, due to the sudden release of energy. But the atom maintains its coherence 
on a longer time scale $T_2\simeq L_K$ due to its long-range spatial entanglement 
with the waveguide, since $\big<\sigma_z(t)\big>$ does not decay.
From the no-cloning theorem~\cite{Sargent}, the quantum information stored in
the dressed atom state cannot be transfered to the wavepacket for times $t\ll T_2$, and 
thus the coherence of the emitted cat state is only partial at intermediate timescales.  
This explains why the negative lobe of the cat state in the right panel
of Fig.~\ref{Wigner} does not quite reach the maximal value $-2/\pi$.

We have to point out that a definite phase is seen in the cat state of Fig.~\ref{Wigner},
which evolves as the state propagates between the emitter and the 
measurement setup. Its absolute value is tied to the original form of the
Hamiltonian~(\ref{hamiltonian}), where the qubit is seen to couple to the first
quadrature of the field. Moreover, it must be stressed that a true measurement
setup will take place outside the waveguide, which must be adapted to a low
impedance environment. It may be that phase information is ultimely lost in
the final output field, as pointed out previously in the context of anisotropic
dielectrics~\cite{Glauber}, hence perhaps modifying the cat structure. 
Interestingly, this problem bears some similarity to transport in interacting quantum wires
(Luttinger Liquids~\cite{Giam}), which can be described by squeezed plasmonic
modes. The interaction fingerprints visible in the conductance of an infinite 
wire are indeed suppressed once the wire is smoothly connected to non-interacting 
leads~\cite{MaslovStone, SafiSchulz}. The question of impedance matching in
ultra-strongly coupled waveguides has not been addressed to our knowledge in the
recent circuit-QED literature~\cite{LeHur,Goldstein,Sanchez,Peropadre,Gheeraert}.

\begin{figure}[th]
\includegraphics[width=0.99\linewidth]{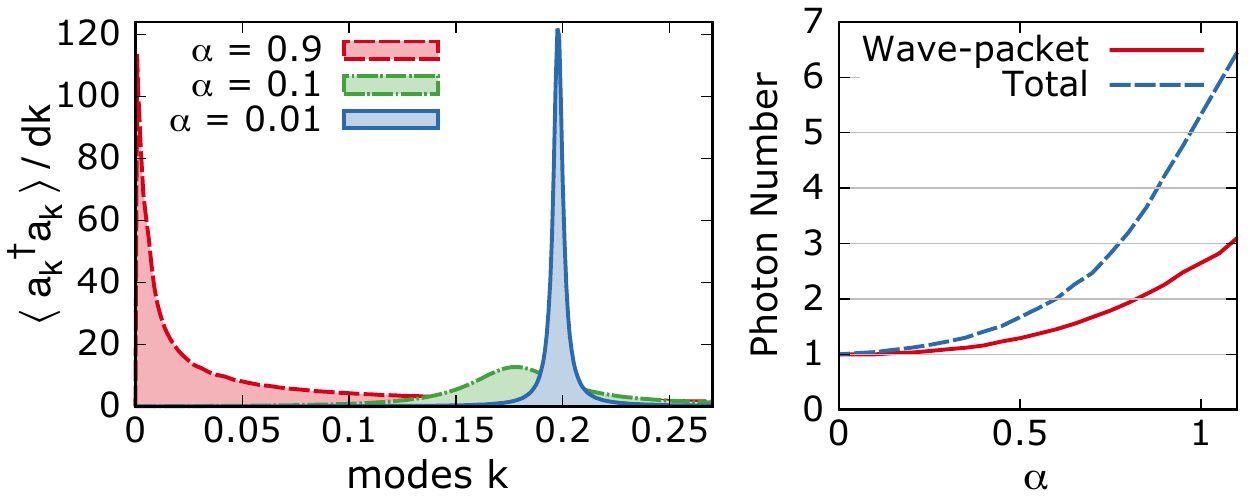}
\caption{(Color online) Left: Emission spectrum showing photon number density $\big<a^\dagger_k
a_k\big>/dk$ in the wavepacket as a function of mode number $k$, for $\alpha=0.01,0.1,0.9$, with 
$\Delta/\w_p=0.2$. Right: integrated photon number in the wavepacket (solid line) and 
in the total wavefunction (dashed line).}
\label{PhotonNumber}
\end{figure}

We conclude our analysis of the emitted cat by displaying its actual photon 
content. In the left panel of Fig.~\ref{PhotonNumber}, we show the spectrally 
resolved photon number density, 
which is obtained as previously by cutting out the static part of the field 
tied to the atom.
In the quantum optics regime $\alpha=0.01$, the lineshape is a narrow Lorentzian 
centered at the bare transition frequency $\Delta$. For increasing $\alpha$, 
the peak is shifted to lower frequencies due to the dressing of the atom 
by the bosonic bath, but more remarkably, its lineshape becomes 
spectrally very broad. This illustrates again the fact that the cat 
state generated by spontaneous emission has a rather localized character in 
the time domain. 
We stress that the number of emitted photons has to vanish
for zero frequency modes, because the coupling constant $g_k$ 
vanishes at $k\to0$. This is also true for the curve $\alpha=0.9$, although
the downturn of the curve can barely be seen on the graph, due to the strong
renormalization of the emission frequency to tiny frequencies.
The size of the cat can be measured by integrating the photon
density curves, giving the photon numbers shown in the right panel of
Fig.~\ref{PhotonNumber}, both for the wavepacket (solid line) and the total
wavefunction (dashed line) comprising wavepacket and entanglement cloud.
At ultrastrong coupling for $\alpha=1.0$, the wavepacket contains $\bar
n_\mathrm{wp}\simeq3$ photons, in agreement with the cat size in
Fig.~\ref{Wigner}, with an average total photon number of $\bar n_\mathrm{tot}=
\bar n_\mathrm{wp}+\bar n_\mathrm{cl}=6$, which highlights the complexity of the 
quantum problem at play, involving more than $1000^{\bar n_\mathrm{tot}}
\simeq 10^{18}$ quantum states.

To summarize, we have found that Schr\"odinger cat states are spontaneously
radiated by a single emitter in an infinite high impedance medium at ultrastrong coupling.
These cats show unusual properties in comparison to standard quantum optics 
protocols~\cite{Haroche,Grangier}: they are spectrally very broad and 
partially quantum coherent at intermediate stages of the dynamics, due to a 
strong separation in time scales between the slow decoherence and the fast 
energy relaxation of the emitter.
After complete information loss of an arbitrary initial state of the qubit
$[u |g\big>+ v |e\big>]\otimes|0\big>$, the qubit relaxes to a unique many-body
ground state $|\mrm{GS}\big>$, associated to the polarization in the waveguide. 
Quantum information is then preserved in a generic cat state
$|\mrm{GS}\big>\otimes[(u+v)|+f\big> + (u-v)|-f\big>]$ of
the radiated field at long times. However, impedance matching of the waveguide to 
a low-impedance external measurement apparatus may drastically affect the quantum 
correlations of spontaneous emission (as well as the structure of inelastically
scattered light~\cite{Goldstein,Gheeraert}), a generic problem that will be 
addressed in a future study.

\section*{Acknowledgments}
We thank H. Baranger, A. Chin, T. Grall, T. Meunier, A. Nazir, 
J. Puertas-Martinez, M. Schir\'o, and I. Snyman for useful discussions, and especially 
N. Roch for stimulating several aspects of this work. Financial support from 
the Nanoscience Foundation is also recognized.

\section*{Appendix: derivation of the dynamical equations}

The starting point of the method is to represent the state vector at an
arbitrary time by an expansion onto coherent states:
\begin{eqnarray}
\label{AppPsi}
|\Psi(t)\ra &=& \sum_{n=1}^{N_\mathrm{cs}} \left[ p_n(t) |f_n(t) \ra | \ua \ra + q_n(t) |h_n(t) \ra | \da \ra 
\right], \\
|f_m(t)\big> &=& e^{\sum_{k=1}^{N_\mathrm{modes}}
[f_m(k,t) \akd - f_m^*(k,t) \ak]} |0\big>,
\end{eqnarray}
where the set of variables $v=\{p_n,q_n,f_n(k),h_n(k)\}$ are complex and time
dependent, with $n=1\ldots N_\mathrm{cs}$. Following the time-dependent variational
principle~\cite{Saraceno}, we define the real Lagrangian density
$\mathcal{L}= \big<\Psi(t)|\frac{i}{2} \overrightarrow{\partial_t} 
-\frac{i}{2} \overleftarrow{\partial_t} - \mathcal{H}|\Psi(t)\big>$,
from which arbitrary variations of the state vector give the simple
Euler-Lagrange equations $\frac{d}{dt} \frac{\partial \mathcal{L}}{\partial \dot{v}} =
\frac{\partial \mathcal{L}}{\partial v}$ for our set of variables. 
For the spin-boson model~\cite{Leggett}, the equations explicitly read:
\begin{eqnarray}
\label{pevolution}
-\ii \f{\partial E}{\partial \conj{p_j}} &=& 
\f{1}{2} \sum_m \Le( 2 \dot{p}_m - p_m \kappa_{mj} \Ri) \la f_j | f_m \ra ,
\\
\nonumber
- \ii \f{\partial E}{\partial \conjk{f_j}} &=&
\sum_m \Big[p_m \conj{p_j} \dot{f}_m^k \la f_j | f_m \ra
  - \f{1}{4}  \Le(2 \dot{p}_m - p_m \kappa_{mj} \Ri) \conj{p_j} (f_j^k-2f_m^k)
\la f_j | f_m \ra \\
\label{fevolution}
&& + \f{1}{4} \Le(2 \conj{\dot{p}_m} - \conj{p_m}\conj{\kappa_{mj}} \Ri) p_j
f_j^k \la f_m | f_j \ra 
\Big] ,\\
\label{kappa}
\kappa_{mj} &=& \sum_{k'>0} [\dot{f}_m^{k'} f^{k' *}_m + \dot{f}_m^{k' *} f^{k'}_m 
- 2 f_j^{k' *}\dot{f}_m^{k'} ].
\end{eqnarray}
Identical equations (up to a minus sign) are obtained for the variables $q_n$ and $h^k_n$. We 
have denoted here $E= \la \Psi | H | \Psi \ra$ the average energy, whose 
explicit expression is:
\begin{align}
E = &\f{\Delta}{2} \sum_{n,m} \Big(\conj{p_n} q_m \la f_n | h_m \ra + p_m
\conj{q_n} \la h_n| f_m \ra\Big) + \sum_{n,m} \Big(\conj{p_n}p_m \la f_n | f_m \ra
W_{nm}^f + \conj{q_n} q_m \la h_n | h_m \ra W_{nm}^h \Big) \nn \\
&- \f{1}{2}\sum_{n,m} \Big(\conj{p_n}p_m \la f_n | f_m \ra L_{nm}^f
- 
 \conj{q_n} q_m \la h_n | h_m \ra L_{nm}^h
 \Big) 
\end{align}
where we have defined $W_{nm}^f=\sum_{k>0} \omega_k f_n^{k*} f_m^k$, 
$W_{nm}^h=\sum_{k>0} \omega_k h_n^{k*} h_m^k$, 
$L_{nm}^{f}=\sum_{k>0} g_k (f_n^{k*} + f_m^k)$, 
$L_{nm}^{h}=\sum_{k>0} g_k (h_n^{k*} + h_m^k)$. 

In contrast to the dynamics with a single coherent state~\cite{Zhao,Bera3},
one encounters here a computational difficulty~\cite{Burghardt}, because 
time-derivatives $\dot{f}_m^{k'}$ of all possible coherent state amplitudes enter 
the dynamical equation ruling a given field $f_j^k$ in Eq.~(\ref{fevolution}) 
through the parameter $\kappa_{mj}$ in Eq.~(\ref{kappa}).
Indeed, for stability reasons it is crucial to formulate the dynamical 
equations in an explicit form $\dot{f}_j^k = F[v]$, where $F$ is only a 
functional of the variables $v=\{p_n,q_n,f^k_n,h^k_n\}$ without reference 
to their time derivatives. Numerical inversion in order to bring the system 
into explicit form is however prohibitive (unless the number of modes is small, 
for instance in the case of the Wilson discretization~\cite{ZhaoMulti}, which 
is not adapted to study the bath dynamics), as it would cost 
$(N_\mathrm{modes} \times N_\mathrm{cs})^3$ operations.
It turns out that a very convenient trick allows to make the inversion in
$(N_\mathrm{cs})^6$ operations, which is favorable provided 
$N_\mathrm{cs}\ll N_\mathrm{modes}$, as is the case for very
long Josephson arrays or broadband environments.
First, we invert Eq.~(\ref{pevolution}-\ref{fevolution}) by 
expliciting the $\kappa_{nj}$ dependence:
\begin{equation}
\label{newevolution}
\dot p_i = \sum_{jn}A_{ijn}\kappa_{nj} + G_i, \;\;\;\;
\dot f_i^s = \sum_{jn} B_{ijns} \kappa_{nj} + H_{is},
\end{equation}
defining the compact notation 
$A_{ijn} = \f{1}{2} M^{-1}_{ij} p_n \braket{f_j | f_n} $, 
$G_i = \sum_j (M^{-1}_{ij} P_j)$, 
$B_{ijns} = \f{1}{2} (N^{-1})_{ij} p_n f_n^s \braket{f_j | f_n} 
- \sum_{lm} (N^{-1})_{il} A_{mjn} f_m^s \braket{f_l | f_m}$, and
$H_{is} = \sum_j (N^{-1})_{ij} F_j^s - \sum_{jm} (N^{-1})_{ij} G_m 
f_m^s \braket{f_j | f_m}$, with $P_j = -i \partial E/\partial p_j^*$, 
$F_j^k = -i\partial E/\partial f_k^{j*}+(1/2)[P_j p_j^*+P_j^* p_j]f_k^{j}$, 
and the overlap matrices
$M_{jm} = \braket{f_j | f_m}$ and
$N_{jm} = p_m \braket{f_j | f_m}$.
The evolution equation~(\ref{newevolution}) is
still not in explicit form, because the parameters $\kappa_{nj}$ explicitly
depend on time derivatives in Eq.~(\ref{kappa}). However, we can now use 
Eq.~(\ref{newevolution}) to replace all the $\dot f_i^k$ terms in
Eq.~(\ref{kappa}), which gives a closed equation for the $\kappa$ matrix:
\begin{eqnarray}
\nonumber
\kappa_{im} &=& \sum_{jns} \Bigl[ (f_i^{s*} - 2 f_m^{s*}) B_{ijns} \kappa_{nj} 
+ f_i^s B_{ijns}^* \kappa_{nj}^* \Bigr] \\
&&+ \sum_s \Bigl[ f_i^{s*} H_{is} + f_i^s H_{is}^* -2 f_m^{s*} H_{is} \Bigr].
\end{eqnarray}
Inverting this linear systems with $(N_\mathrm{cs})^2$ parameters provides
the final $(N_\mathrm{cs})^6$ scaling of our algorithm. Of course the gain is
important only provided that $N_\mathrm{cs}$ stays small during the unitary time
evolution. We show now that the dynamics indeed converges rapidly for a 
surprisingly small number of coherent states, demonstrating that the 
decomposition~(\ref{AppPsi}) is physically well motivated.

\end{document}